# Quantum Nonlocality: how does Nature do it?

Marian Kupczynski

Département de l'Ingénierie, Université du Québec en Outaouais (UQO), Case postale 1250, succursale Hull, Gatineau. QC, Canada J8X 3X7
* Correspondence: marian.kupczynski@uqo.ca ORCID: MK, 0000-0002-0247-7289

**Abstract:** In his article in Science, Nicolas Gisin claimed that quantum correlations emerge from outside space–time. We explain that they are due to space-time symmetries. This paper is a critical review of metaphysical conclusions found in many recent articles. It advocates the importance of *contextuality*, Einstein-causality and global symmetries. Bell tests allow only rejecting probabilistic coupling provided by a local hidden variable model, but they do not justify metaphysical speculations about quantum nonlocality and objects which know about each other's state, even when separated by large distances. The violation of Bell inequalities in physics and in cognitive science can be explained using the notion of Bohr-*contextuality*. If contextual variables, describing varying experimental contexts, are correctly incorporated into a probabilistic model, then the Bell–CHSH inequalities cannot be proven and nonlocal correlations may be explained in an intuitive way. We also elucidate the meaning of *statistical independence* assumption incorrectly called *free choice*, *measurement independence* or *no- conspiracy*. Since correlation does not imply causation, the violation of *statistical independence* should be called *contextuality*. It does not restrict the experimenter's freedom of choice thus, contrary to what is believed, closing the *freedom-of choice loophole* does not close the *contextuality loophole*.

1. Introduction

Since many years, physical community has been puzzled by a paradox: how <u>outcomes</u> of distant measurements (<u>created</u>, according <u>to some interpretations</u> of quantum mechanics (QM), <u>in a perfectly random way</u>) <u>could be perfectly correlated</u> as it was predicted for an ideal EPRB experiment [1]. In 1936, Einstein pointed out in [2], that to avoid this paradox, one should reject a genuine randomness and adopt a statistical interpretation (SI) of QM according to which a quantum state/wave function describes only an ensemble of identically prepared physical systems. In SI, individual physical systems are described, as in classical physics, by unknown but precise values of some observables which are measured with errors by macroscopic instruments. Therefore, perfectly correlated outcomes of measurements in an ideal EPRB experiment could be explained by strictly correlated properties of "entangled pairs" prepared by a source. The predetermination of experimental outcomes by some ontic properties of physical systems, coded by hidden variables independent on measuring instruments, has been called *local realism* or *counterfactual definiteness* (CFD).

As Bell [3] and Clauser-Horne-Shimony-Holt [4] demonstrated, a local realistic hidden variable model (LRHVM), inspired by CFD, cannot reproduce quantum predictions for an ideal EPRB experiment. Ideal EPRB experiments do not exist [5, 6] . Nevertheless, several experiments have been performed allowing to compare quantum predictions with the predictions obtained using LRHVM. According to LRHVM, a probability that some cyclic combinations of pairwise expectations violate Bell-CHSH inequalities is tending to 0, if sample sizes are increasing to infinity. Since a significant violation of the inequalities, was reported in several loophole free Bell Tests [7-17] one might conclude, that neither LRHVM nor stochastic hidden variable models (SHVMs) [18] provided a correct description of these experiments.

Since the violation of Bell-CHSH inequalities is a source of many extraordinary metaphysical conclusions and speculations, we discuss below in some detail the meaning and implications of several recent Bell Tests. The interpretation of quantum experiments and phenomena may depend on the interpretation of QM adopted by authors. For us, a probability is an objective property of random experiments and QM is a theory providing precise probabilistic predictions for random experiments performed, in well-defined reproducible experimental contexts, on ensembles of identically prepared physical systems or repeated experiments performed on the same physical system [19-23].

This article is our answer to the question of Nicolas Gisin: Quantum nonlocality: how does nature do it? He concluded that quantum correlations emerge from outside space-time [24]. We will explain that they are due to *contextuality* and global symmetries of space-time.

The paper is organized as follows. In section 2, we compare classical and quantum observables and discuss shortly joint probability distributions and non-contextuality inequalities. In section 3, we introduce Bell Tests. In section 4, we discuss empirical probability distributions and various probabilistic couplings tested in Bell Tests.

In section 5, we define and analyze in detail a local hidden variable model rejected in the recent loophole free Bell Tests. In section 6, we define a contextual hidden variable model, allowing to explain the violation of Bell-CHSH inequalities and "nonlocal" correlations. In section 7, we discuss the *detection loophole* in some earlier photonic Bell Tests and explain in some detail an apparent violation of non-signalling reported in [8]. In section 8, we describe shortly non-photonic tests immune to the *detection loophole*. The section 9 contains our conclusions. In the Appendix, we present in some detail experimental protocols used in the most recent Bell Tests [10, 15-17].

## 2. Classical properties versus quantum observables

In classical physics, the properties of physical systems are quantified by values of various compatible observables, which can be measured in any order. Measurement outcomes may contain experimental errors but measurements are assumed to be non-invasive, what means that they do not change the properties they measure.

In quantum experiments, the information obtained about invisible physical systems is indirect and obtained from their interactions with different macroscopic measuring instruments. As Bohr correctly insisted, the atomic phenomena are characterized *"the impossibility of any sharp separation between the behaviour of atomic objects and the interaction with the measuring instruments which serve to define the conditions under which the phenomena appear"* (Bohr ([25], v. 2, pp. 40–41). Quantum observables have the following properties [26]:

*Bohr-contextuality: The output of any quantum observable is indivisibly composed of the contributions of the system and the measurement apparatus.*

*Bohr-complementarity: There exist incompatible observables (complementary experimental contexts)* [27].

*KS-contextuality: A measurement of an observable does not need to yield the same value independently of what other measurements are made simultaneously* [28].

Inspired by KS-contextuality, Dzhafarov and Kujala proposed Contextuality-by-Default (CbD) approach in which random variables describing outcomes of measurements, performed in physics and in other domains of science, are labelled not only by a measured content but also by an experimental context. CbD approach found a lot of applications and is investigated by several scientists [29-36].

In mathematical statistics, multivariate random variables and joint probability distributions are only used to describe random experiments or population surveys, in which each trial/individual is described not by one, but by several data items. In this case we say that these data items are 'measured' values of commeasurable/compatible random variables [36]. Einstein believed that quantum pure ensembles are in fact mixed statistical ensembles of physical systems, which may be described by joint probability distributions of non-contextual random variables (NCHV). Cyclic combinations of pairwise marginal expectations of binary random variables must obey some noncontextuality inequalities (NCIs) [37]. Bell-CHSH inequalities are a special case of NCI and they are significantly violated by the experimental data in Bell Tests.

### 3. Bell Tests

Bell Tests are subtle and complicated experiments run according to several different protocols. They are inspired by an ideal EPRB experiment in which specific correlated measurements are performed, using 4 randomly chosen settings, on ensembles of prepared correlated/entangled pairs of physical systems. Entangled pairs may be created at the source and sent to distant locations, as it was done in many Bell Tests [7-9, 11-14], or they are created directly in distant laboratories using specific synchronized preparations/treatments, as it was done in experiments [10,14-17] using an *entanglement swapping* or *entanglement transfer* protocols In spite of differences, experimental protocols are subdivided into 3 steps:

(1) <u>Preparation of an ensemble E</u> of pairs of entangled physical systems.
(2) <u>Random local choice of labels (x, y)</u> (called inputs) of 4 incompatible experimental settings using random number generators (RNG), signals coming from the distant stars [ 13] or /and  human choices [14-17]. In this article we use 4 pairs of labels:    (x, y), (x, y'), (x', y) and (x', y'), which denote 4 different experimental contexts.
(3) <u>Implementation of correlated and synchronized measurements</u> in distant locations and readout of binary outcomes (a, b) (called outputs), which are the coded information corresponding to clicks on different distant detectors etc.

Therefore, in a Bell Test one performed 2 local independent random experiments to choose 4 pairs of labels and 4 pairs of correlated distant random experiments corresponding to these 4 choices. In several papers, one uses a different notation: (a, b) to denote inputs and (x, y) to denote outputs.

### 4. Empirical probabilistic models and probabilistic couplings

In mathematical statistics and in quantum mechanics random experiments are described as "black boxes".  Outcomes of a random experiment, in which empirical frequencies stabilize, are the values of some random variable obeying a specific probability distribution. Therefore, experimental outcomes in Bell Tests are described a priori by 8 binary random variables: ($A_{xy}$, $B_{xy}$), ($A_{xy'}$, $B_{xy'}$), ($A_{x'y}$, $B_{x'y}$) and ($A_{x'y'}$, $B_{x'y'}$) [36]. We are using a notation inspired by the CbD approach  in which random variables measuring the same content in a different context are stochastically unrelated e.g. $A_{xy}$ and $A_{x'y}$. It is evident, that a joint probability distribution of these 8 random variables does not exist, thus CHSH inequality cannot be derived without additional assumptions [36, 38].

A pair of random empirical variables ($A_{xy}$, $B_{xy}$) describes a scatter of outputs in the experiment using settings (x, y) .  In a Bell Test we have 4 random experiments described by 4 specific probabilistic empirical models. We want to explain more in detail how outputs might have been created and how their scatter depends on chosen settings and the preparation of a particular ensemble E of physical systems. This is why several hidden variable probabilistic models were created and their plausibility tested.

In this article, random variables in probabilistic models are denoted (A′$_{xy}$, B′$_{xy}$) in order to be not confounded with empirical random variables (A$_{xy}$, B$_{xy}$). In Bell Tests, we check only a plausibility of different probabilistic couplings. We say that a probabilistic model provides a probabilistic coupling if:

$$E(A_{xy}) = E(A'_{xy}), E(B_{xy}) = E(B'_{xy}), E(A_{xy}B_{xy}) = E(A'_{xy}B'_{xy}) \quad (1)$$

i) Quantum coupling

QM does not give any details how outputs are created and provides only specific probabilistic predictions; e.g. if the settings (x, y) are chosen then $E(A'_{xy}B'_{xy}) = Tr(\rho \hat{A}_x \hat{B}_y)$, where ϱ is a density matrix describing the ensemble E $\hat{A}_x$ and $\hat{B}_y$ are Hermitian operators representing synchronized measurements made by Alice and Bob in distant laboratories.

In an ideal EPRB experiment the ensemble E is described by $\rho = |\psi\rangle\langle\psi|$, $\hat{A}_x = \vec{\sigma} \cdot \vec{n}_x$ and $\hat{B}_y = \vec{\sigma} \cdot \vec{n}_y$ represent spin projections on the corresponding unit vectors thus:

$$E(A'_{xy}B'_{xy}) = \langle\psi | \hat{A}_x \otimes \hat{B}_y | \psi\rangle = \sum_{\alpha\beta} \alpha\beta p_{xy}(\alpha,\beta) = -\vec{n}_x \cdot \vec{n}_y = -\cos(\theta_{xy}) \quad (2)$$

where $\hat{A}_x \otimes \hat{B}_y |\alpha\beta\rangle_{xy} = \alpha\beta |\alpha\beta\rangle_{xy}$, $p_{xy}(\alpha,\beta) = |\langle\psi|\alpha\beta\rangle_{xy}|^2$ and α=±1 and β=±1 [39].

In QM, each incompatible experimental context is described by a specific dedicated probabilistic model $\{\rho, \hat{A}_x \otimes \hat{B}_y\}$ defining a *quantum probabilistic coupling* (2).

ii) Local realistic coupling (LRHVM)

For x=y, QM predicts strictly anti-correlated outcomes for all the directions $\vec{n}_x$, what could not be true, if the outcomes were produced in an irreducible random way, as it was believed. Therefore, Bell [40, 41] concluded that outcomes had to be predetermined at the source and proposed LRHVM, <u>which should be understood as a probabilistic coupling</u> [38]:

$$E(A'_{xy}B'_{xy}) = E(A'_x B'_y) = \sum_{\lambda \in \Lambda} A'_x(\lambda) B'_y(\lambda) P(\lambda) \quad (3)$$

where $A'_x(\lambda) = \pm 1$ and $B'_y(\lambda) = \pm 1$. In LRHVM, we have 4 jointly distributed random variables (A′$_x$(L), B′$_y$(L), A′$_{x'}$(L), B′$_{y'}$(L)), being deterministic functions of a hidden random variable L. From (3) one can easily derive CHSH inequality:

$$|S| = |E(A'_x B'_y) + E(A'_x B'_{y'}) + E(A'_{x'} B'_y) - E(A'_{x'} B'_{y'})| \leq 2. \quad (4)$$

The random variable L is describing a classical random experiment in which λ is sampled with replacement from a probability space Λ, which does not depend on experimental settings (x, y). LRHVM describes *entangled pairs* as pairs of socks, which can have different sizes and colours; e.g. Harry draws a pair of socks, sends one sock to Alice and another to Bob, who in function of (x, y) record corresponding properties colour or size. The summation in (3) and in other equations in this article may be replaced by a suitable integration.

In the discussion of Bell Theorem instead of (3) one is usually defines LRHVM by: $E(A_x B_y) = \sum_{\lambda \in \Lambda} A_x(\lambda) B_y(\lambda) P(\lambda)$ and $E(A'_{xy} B'_{xy}) = E(A_x B_y)$, This is misleading, because (A′$_x$(L), B′$_y$(L), A′$_{x'}$(L), B′$_{y'}$(L)) are jointly distributed in contrast to empirical random variables, denoted <u>imprecisely</u> as ($A_x$, $B_y$, $A_{x'}$, $B_{y'}$). The only way to avoid misunderstanding is to use 8 random variables ($A_{xy}$, $B_{xy}$)…($A_{x'y'}$, $B_{x'y'}$) to describe the experimental data and the concept of the probabilistic coupling [29-31, 36, 38].

The probabilistic coupling defined by LRHVM is constrained by CHSH inequality (4). Bell clearly demonstrated that LRHVM is inconsistent with QM, since there exist 4 particular experimental settings for which using (2) one obtains: $S = 2\sqrt{2}$, which significantly violates (4). As Boris Tsirelson [42] and Lev Landau [43] demonstrated, this is the maximal value of S allowed for any density matrix $\rho$ and any Hermitian operators $\|\hat{A}_x\| \leq 1$ and $\|\hat{B}_y\| \leq 1$. Therefore, <u>the quantum probabilistic coupling</u> (2) is constrained by the Tsirelson inequality, called by Andrei Khrennikov quantum- CHSH inequality [6, 44].

    iii)    Local stochastic coupling (SHVM)

An ideal EPRB experiment does not exist. Nevertheless, LRHVM may be tested for example in spin polarization correlations experiments (SPCE), in which a physical system decays from an initial state J=0 to an intermediate state J=1 and then reaches the J=0 ground state by emitting a pair photons. In these experiments there are no perfect correlations, thus Clauser and Horne [18] proposed SHVM in which λ does not determine outcomes in a given trial but only their probability. Using the notation of Big Bell Test collaboration SHVM may be defined as follows:

$$P(a,b|x,y) = \sum_\lambda P(a|x,\lambda) P(b|y,\lambda) P(\lambda) \quad (5)$$

where P(-|-) denotes a conditional probability. The equation (5), for a fixed setting (x, y) describes a family of independent random experiments labelled by λ and:

$$E(A'_{xy} B'_{xy}) = \sum_\lambda E(A|x,\lambda) E(B|y,\lambda) P(\lambda) = P(a=b|x,y) - P(a \neq b|x,y) \quad (6)$$

In [18], it was still assumed, that λ were old fashioned local hidden variables representing ontic properties of entangled pairs, thus entangled photon pairs were described as pairs of dice and the correlations, which might had been created in this way, were quite limited. Nevertheless, if a density matrix ϱ is a mixture of separable quantum states, then SHVM can explain all quantum correlations. We discussed it in detail in [45].

**5. Bell local causality and local hidden variable model (LHVM)**

Nowadays, λ's in (5) represent causes in the past and as some believe: "*they may include the usual quantum state; they may also include all the information about the past of both Alice and Bob. Actually, the λ's may even include the state of the entire universe*" [14], except that inputs (x, y) cannot depend on them.

No matter what metaphysical or causal arguments might motivate a choice of a probabilistic coupling (5), we stay only on safe grounds, if we assume that λ's are values of some multivariate hidden random variable.

The model (3) is a special case of the model (5), if P(a| x, λ) and P(b| y, λ) are 0 or 1, Nevertheless, (3) and (5) describe random experiments using completely different experimental protocols. LRHVM is motivated by *local determinism*, and SHVM by *local causality* as Bell defined it. An extensive discussion of these concepts and of two Bell Theorems was given by Wiseman [46, 47]. Since, both models are inconsistent with quantum mechanics and with the experimental data they may simply be called LHVM (local hidden variable model).

LHVM is not completely defined by (5, 6). Additional equations and conditions are added:

$$P(a, b, x, y) = \sum_{\lambda} P(a | x, \lambda) P(b | y, \lambda) P(x, y | \lambda) P(\lambda) \quad (7)$$

and

$$P(x, y | \lambda) = P(x, y). \quad (8)$$

It is important to point out, that the notation used in (7) and (8) is imprecise and may be misleading. In fact, when we are talking about probabilities we are talking about events and random variables [22], thus P (a, b, x, y) should be understood as a shorthand notation of P (A=a, B=b, X=x, Y=y), where A, B, X and Y are the corresponding random variables. Similarly P(x, y| λ) is a short hand notation of P({x, y}|{λ}), where an event {x, y}= {(X=x, Y=y)} and a hidden event {λ}={L=λ}. Besides only random variables X, Y and L can be stochastically dependent and not their values. Moreover, if one chooses inputs without using specific random experiments, then the equation (8) is meaningless. Noting all that, we will not be pedantic and we will follow bellow the standard shorthand notation and the terminology used in [14] and in nearly all the papers on the subject.

The important condition (8) is called *measurement independence*, *freedom- of- choice* (FoC) or *no conspiracy* [48-51]. This terminology is based on an incorrect causal interpretation of conditional probabilities [22, 38]. Namely, $P(x, y | \lambda) \neq P(x, y)$ *was* believed to constrain experimenter's freedom of choice by some unspecified causal influences from the past.

For Bell, experimental settings (x, y) could be chosen at a whim of experimenters, without talking about probabilities P(x, y). Moreover, $\lambda$'s in (3) described ontic properties of entangled pairs, thus they could not depend in any sense on the chosen settings. Using the notation consistent with (7, 8) it means that:

$$P(\lambda | x, y) = P(\lambda). \tag{9}$$

If $\lambda$ are the values of some random variables, then the equation (9) says only that $\lambda$ does not depend statistically on (inputs), this is why it should be called *statistical independence* (SI) (a terminology adopted by increasing number of authors [52, 53]) or *noncontextuality* [22, 36, 38] and not *free choice* or FoC as many authors continue to call it.

In fact (8) and (9) are two equivalent forms of the condition: $P(\lambda, x, y) = P(\lambda)P(x, y)$. The violations of these condition means only that $(\lambda, x, y)$ are statistically dependent. It does not justify the conclusions: "$P(x, y | \lambda) \neq P(x, y)$ *thus $\lambda$ is a free variable which is a cause of (x, y)*" or "$P(\lambda | x, y) \neq P(\lambda)$ *thus (x, y) are free variables which are causes of $\lambda$*". In some everyday situations and in medicine a causal interpretation of conditional probabilities may be justified, but not as it was used in the discussions of Bell-CHSH inequalities [22].

Already in 1964, Bell was fully aware of the fact that, if hidden variables depended on the settings Bell-CHSH inequalities could not be proven. Unfortunately, Shimony, Horne and Clauser [49, 54] convinced him, that the violation of (9) implied the violation of (8), what interpreted <u>incorrectly</u> in a causal way would mean *superdeterminism*. Thinking that he had to choose between *superdeterminism* or *nonlocality* Bell opted for *nonlocality*. In [55], where (a, b) denote the settings, he summarized his point of view as follows:

> *We supposed ... that a and b could be changed without changing the probability distribution ρ(λ). Now even if we have arranged that a and b are generated by apparently random radioactive devices, housed in separate boxes and thickly shielded, or by Swiss national lottery machines, or by elaborate computer programs, or by apparently free willed experimental physicists, or by some combination of all of these, we cannot be sure that a and b are not significantly influenced by the same factors λ that influence A and B. But this way of arranging quantum mechanical correlations would be even more mind boggling that one in which causal chains go faster than light.*

*Apparently separate parts of the world would be deeply and conspiratorially entangled, and our apparent free will would be entangled with them.*

This conclusion is incorrect. The violation of a *statistical independence* does not need to be due to causal chains going faster than light or the lack of FoC.

There is a lot of confusion in the literature and on the social media concerning metaphysical implications of the results of Bell Tests [14]. Thus, let us first recall what a Bell Test is. Using LHVM (5-9), one derives inequalities which have to be satisfied by specific combinations of probabilities of events to be observed in the experiments performed using different experimental setting. These combinations are denoted S, J or T, which are shortly called Bell parameters. As it is explained clearly in the Methods section [14]:

*A Bell Test is an experiment which makes many spatially-separated measurements with varied settings to obtain estimates of P(a,b|x,y), that appear in a Bell parameter. If the observed parameter violates the inequality, one can conclude that measured systems were not governed by any LHVM. It should be noted that this conclusion is always statistical, and typically takes a form of a hypothesis test, leading to a conclusion of the form: 'assuming nature is governed by local realism, the probability to produce the observed Bell inequality violation ... is P(observed or stronger | local realism )≤ p. This p-value is a key indicator of statistical significance in Bell Tests.*

Since p-values in several experiments are very small one concludes: "Local realism, i.e., <u>realism plus relativistic limits on causation</u>, was debated by Einstein and Bohr using metaphysical arguments, and recently has been rejected by Bell tests" [14]. Such conclusion is imprecise, misleading and has been a source of unfounded speculations about quantum magic.

As Wiseman correctly pointed out in [46]: "*the usual philosophical meaning of "realism" is the belief that entities exist independent of the mind, a worldview one might expect to be foundational for scientists.*" This point of view was also shared by Bell, who was in fact a realist [38, 55, 56]. The *local realism* should be rather called *local determinism* (LD) or *counterfactual definiteness* (CFD) [38, 46] and defined as: results of any measurement on an individual system are predetermined by some ontic properties, which have definite values, whether they are measured or not. Any probabilities we may use to describe the system merely reflect our ignorance of these hidden definite values, which may vary from one experimental run to another.

It is claimed that Bell tests allow to reject not only *local realism* but also *local causality*, where Bell-local causality is defined: Alice's output *a* depends only on her input *x* and on $\lambda$ describing all possible common causes included in the intersection of the of the backward light cones of *a* and *b* and <u>independent on inputs *x* and *y*</u>.

It is true that tested probabilistic models have been motivated by CFD or *Bell-local causality*. Nevertheless, Bell Tests only allow rejecting the statistical hypothesis saying that LHVM (5-9) provides a probabilistic coupling consistent with experimental data. Therefore, the violation of Bell-CHSH inequalities does not allow for the far reaching metaphysical speculations. Let us cite here Hans de Raedt et al .[57]: " *all EPRB experiments which have been performed and may be performed in the future and which only focus on demonstrating a violation BI-CHSH merely provide evidence that not all contributions to the correlations can be reshuffled to form quadruples… These violations do not provide a clue about the nature of the physical processes that produce the data….*" Similar conclusions may be found in [6, 44, 45, 58-62].

Bell-local causal model (5-9) is incomplete, because P($\lambda$) does not depend on changing experimental measurement contexts. If additional context dependent, variables, describing measuring instruments and procedures are correctly incorporated into the probabilistic model (3), then Bell-CHSH inequalities cannot be derived and the "nonlocal " correlations can be explained without evoking quantum magic. We discuss such model in the next section.

## 6. Contextual hidden variable model and statistical independence

We incorporate into the model (3) additional variables describing the distant measuring contexts. Moreover, we split $\Lambda$ into 4 separate subsets and we assume that:

- $\lambda_1 \in \Lambda_1$ and $\lambda_2 \in \Lambda_2$ describe correlated physical systems on which measurements are made in Alice's and Bob's laboratories. They do not depend on measurement contexts (x, y).
- $\mu_x \in M_x$ and $\mu_y \in M_y$ describe measurement procedures and instruments at the moment of measurement, when the settings (x, y) were chosen in Alice's and Bob's distant laboratories. They do not depend on how the systems were prepared.
- Free inputs (x, y) are randomly chosen binary labels of local measurement contexts. They are chosen in separate random experiments independent on the variables describing a preparation of physical systems and the subsequent measurements. Nevertheless $\mu_x$ depends statistically on *x* and $\mu_y$ depends statistically on *y*. We also assume that $\mu_x$ and $\mu_y$ are causally independent but they can be statistically dependent.
- Binary outputs are created locally in a deterministic way: $a = A'_x(\lambda_1, \mu_x) = \pm 1$ and $b = B'_y(\lambda_2, \mu_y) = \pm 1$

The resulting contextual model (CHVM) is defined by three equations

$$E(A'_{xy} B'_{xy}) = \sum_{\lambda \in \Lambda_{xy}} A'_x(\lambda_1, \mu_x) B'_y(\lambda_2, \mu_y) P(\lambda_1, \lambda_2) P_{xy}(\mu_x, \mu_y) \quad (10)$$

where $\Lambda_{xy} = \Lambda_1 \times \Lambda_2 \times M_x \times M_y$,

$$P(a, b, x, y) = \sum_{\lambda} P(a | \lambda_1, \mu_x) P(b | \lambda_2, \mu_y) P(\mu_x, \mu_y | x, y) P(x, y) P(\lambda_1, \lambda_2) \quad (11)$$

and

$$P(\mu_x, \mu_y | x, y) = P_{xy}(\mu_x, \mu_y) \neq P(\mu_x, \mu_y). \quad (12)$$

In Bell Tests, P(x, y) = P(x) P(y), but in the contextual model (10-12) and in QM, it does not matter how the labels (x, y) are chosen.

The model (10-12) violates *statistical independence* and $P(x, y | \mu_x, \mu_y) \neq P(x, y)$:

$$P(\mu_x, \mu_y, x, y) = P_{xy}(\mu_x, \mu_y) P(x, y) = P(\mu_x, \mu_y) \rightarrow P(x, y | \mu_x, \mu_y) = 1 \quad (13)$$

The equation: $P(x, y | \mu_x, \mu_y) = 1$ "tells" only, that if a hidden event $\{\mu_x, \mu_y\}$ 'happened' then the settings (x,y) were used [22, 36, 38]. It has nothing to do with *conspiracy* and FoC.

As we discussed in the preceding section *statistical independence* (9), called *measurement independence* [17, 48] or *free choice* [14. 49-51] was needed to derive Bell-CHSH inequalities. Therefore, it was <u>incorrectly</u> believed that to prove the *statistical independence* it was sufficient to close *fredom- of- choice loophole* (FoCL). The inputs (x, y) were chosen using signals coming from distant stars [13], random number generators or using random human choices made during online computer games [14], thus they could not causally depend on any variables describing subsequent measurements.

We have no doubt that, in Bell Tests [8-17, 63], FoCL was successfully closed, but it did not prove that hidden variables could not depend on experimental contexts. In CHVM *experimenters' freedom of choice* is not compromised but *statistical independence* is violated. This is why *the violation of statistical independence* should be called: *Bohr-contextuality* or simply *contextuality* not to be confounded with CbD-contextuality [29-35]. As Theo Nieuwenhuizen explained several years ago the *contextuality loophole* (the violation of (9)) is a theoretical loophole and it can never be closed [64,65].

<u>CHVM is neither local nor nonlocal</u>. The inputs and outputs are created locally but variables describing physical systems and measuring contexts, in space-like separated laboratories, can be statistically correlated. As we explain below this correlation may be explained without evoking spooky influences. It may be the effect of setting dependent post-selection of data or [4,5,36] or it may due to the global space-time symmetries [38].

Several loopholes [66] had been successfully closed in Bell Tests, but an ideal EPRB experiment does not exist. In an ideal EPRB thought experiments in each setting (x,y) we have a steady flow of twin-electron or twin-photon pairs producing correlated clicks on distant detectors coded by values of random variables ($A_x$, $B_y$), where $A_x = \pm 1$ and $B_y = \pm 1$. There are no losses of pairs and all expectations $E(A_x B_y)$ may be <u>unambiguously</u> estimated using experimental data [5, 36]. Moreover, one has to assume perfect reproducibility of the ensemble of pairs E, which does not depend on the settings (x, y) [60].

The experimental situation in Bell Tests is much more complicated. One has to construct samples of correlated clicks or other events which may be interpreted as a results of measurements performed on pairs of entangled physical systems prepared by a distant source or by using so called entanglement swapping or entanglement transfer protocols.

In order to find out how significantly the inequalities are violated, we have to assume fair sampling from the same statistical population of the physical systems on which subsequent measurements are performed in different experimental settings.

In the next two sections we discuss experimental challenges and reported anomalies in two different types of Bell Tests.

## 7. Photon identification loophole, data post-selection and anomalies

In spin polarisation experiments (SPCE) correlated photonic signals are sent from a source to Alice's and Bob's measuring stations. Before signals arrive, measurement settings (x,y) are chosen and each detected click is coded +/-1 and outputted together with its time tag. These two time series, are converted using synchronized time-windows of width $W$ into raw data containing correlated pairs ($a_r$, $b_r$), where $a_r$ and $b_r$ are 0 or +/-1. From the raw data only pairs of non-zero outputs are extracted and expectations $E(A_{xy} B_{xy})$ are estimated. Four estimated pairwise expectations are used to determine a significance of the violation of CHSH inequality (4). All steps described above can be hidden in coincidence circuitry outputting only the final data [5, 36].

The interpretation of reported violation of CHSH is not unambiguous because in each SPCE, there are black counts, laser intensity drifts, photon registration time delays etc. Besides , it is important to check <u>carefully</u> that trials are independent and identically distributed. We demonstrated with Hans de Raedt [67,68] that without such verification the standard statistical inference is not reliable. A detailed discussion of experimental protocols and possible loopholes in Bell Tests may be found in Larsson [66].

However, the most troubling problem was, (a much more detailed discussion may be found in [5, 36]) a significant violation of parameter independence/no-signalling. It was discovered and discussed by several authors :Adenier and Khrennikov [69,70], De

Raedt, Jin and Michielsen [71, 72], Bednorz [73], Liang and Zhang [74].

This apparent violation of no-signalling did not mean that the Einsteinian no-signalling had been violated, because raw single count data were free from this anomaly [5]. In series of papers we pointed out that the raw and the final data in Weihs et al. experiment [8] could be described, without evoking *spooky influences*, using a partcular contextual probabilistic model [5, 6, 22, 36] in which the violation of inequalities and of no-signalling was due to *contextuality* and setting dependent post-selection.

In this model the instrument variables were denoted: $\lambda_x \in \Lambda_x$ and $\lambda_y \in \Lambda_y$

$$E(A'_{xy} B'_{xy}) = E(A_x B_y \mid A_x B_y \neq 0) = \sum_{\lambda \in \Lambda'_{xy}} A_x(\lambda_1, \lambda_x) B_y(\lambda_2, \lambda_y) P(\lambda) / C_{xy} \quad (14)$$

$$E(A'_{xy}) = E(A_x \mid A_x B_y \neq 0) = \sum_{\lambda \in \Lambda'_{xy}} A_x(\lambda_1, \lambda_x) P(\lambda) / C_{xy} \quad (15)$$

$$E(B'_{xy}) = E(B_y \mid A_x B_y \neq 0) = \sum_{\lambda \in \Lambda'_{xy}} B_y(\lambda_2, \lambda_y) P(\lambda) / C_{xy} \quad (16)$$

where $A_x(\lambda_1, \lambda_x) = 0, \pm 1$, $B_y(\lambda_1, \lambda_x) = 0, \pm 1$,
$P(\lambda) = P_x(\lambda_x) P_y(\lambda_y) P(\lambda_1, \lambda_2)$,
$\Lambda'_{xy} = \{\lambda \in \Lambda_{xy} \mid A_x(\lambda_1, \lambda_x) \neq 0, B_y(\lambda_2, \lambda_y) \neq 0\}$ and
$C_{xy} = P(A_x B_y \neq 0)$.

To describe the raw data we assumed $P_{xy}(\lambda_x, \lambda_y) = P_x(\lambda_x) P_y(\lambda_y)$ and we derived the model (14-16) describing the final data. The model (13-14) may be rephrased as a special case of a general CHVM in which $P_{xy}(\lambda_x, \lambda_y)$ do not factorize [38]. In the papers [5, 6, 22, 36], the constants $C_{xy}$, were missing but it did not change any conclusions because the equations (14-16), were never used to make quantitative predictions. We are indebted to Richard Gill for noticing it. The setting labels (x, y) were never replaced by numerical values used in the experiments in order to avoid a possible confusion of $(\lambda_x, \lambda_y)$ with $(\lambda_1, \lambda_2)$. To avoid such confusion in future, in CHVM we replaced $(\lambda_x, \lambda_y)$ by $(\mu_x, \mu_y)$.

In contrast to QM and LHVM, the model discussed in this section can explain a significant violation of Bell inequalities in the experiment of Iannuzzi et al. [75], in which the inequalities were violated in spite of independent sources of polarized photons.

In our model probabilistic spaces $\Lambda_{xy}$, describing experiments performed in different settings do not overlap, thus, as Larsson and Gill [76] demonstrated, $S$ is only bounded by 4. In the experiment

[8], the *detection loophole* [77-80] was not closed because in order to test Bell-CHSH inequalities one had to extract/post-select non-vanishing pairs of outcomes [5]. In our model (14-16), such post-selection is explicitly incorporated. When discussing photonic Bell Tests it is more appropriate to talk about *photon- identification loophole* [81, 82] and not about *detection loophole*, because the latter term tells only that detectors do not function as they should.

Already in 1970, Philippe Pearle constructed and discussed in detail local hidden variable models based on the data rejection able to reproduce quantum predictions [77]. This is why in the most recent experiments a considerable efforts were made to close the *detection loophole*.

## 8. Recent loophole free experiments based on the entanglement transfer.

In the classical EPRB-type experimental protocol "entangled pairs" are produced by a source and sent to distant measuring stations A and B [4, 5, 7-9, 11-14]. The recent experiments [10, 15, 17] use different physical qubits and different experimental protocols based on the *entanglement transfer* and *entanglement swapping* [83-85].

The qubits in the distant nodes A and B are trapped physical systems or artificial atoms which may be in a ground or excited state. They oscillate between these two states under the influence of laser pulses. Passing from the excited to ground states they emit photons which carry a local information about qubits. These photons are used to create an entanglement between distant qubits.

When a successful entanglement is created, inputs (x, y) (being the labels of rotation angles $(\theta_x, \theta_y)$) are randomly chosen and qubits are rotated locally using corresponding laser pulses or microwaves. Next, "states" of qubits are locally "measured" . If the qubit in the node A is found in a ground state and the qubit in the node B is found in an excited state, then the outputted readout is: (a ,b)=(1,-1) etc.

Next the distant qubit are reset and all the steps are repeated. After repeating these steps many times one can estimate pair-wise expectations E($A_{xy}B_{xy}$), test CHSH inequality and study how E($A_{xy}B_{xy}$) oscillate in function of the angle $\theta_{xy} = \theta_x - \theta_y$.

Since each trial produces a valid read-out, *detection loophole* is definitely closed.

The detailed discussion of the *entanglement swapping* and *entanglement transfer* protocols is beyond the scope on this article, but we give some more information about Delft, Munich and Zurich experiments in the Appendix.

## 9. Conclusions

The significant violation of Bell-CHSH inequalities was confirmed in the Bell Tests discussed above and in several experiments in the cognitive science [32, 86-90]. As Dirk Aerts demonstrated many years ago [91], these inequalities can be easily violated in simple macroscopic experiments.

In many articles published in high impact journals one may find conclusions that the violation of Bell-CHSH inequalities allowed to reject with a great confidence *local realism* and *local causality*. These conclusions taken out of the context lead to extraordinary metaphysical speculations in the social media, in the papers and books addressed to a general public.

It is true, that local hidden variable models had been motivated by various metaphysical assumptions. Unfortunately, the discussions of possible causal explanation of quantum correlations in [14, 46, 49, 50, 54, 92-95], may lead to unfounded metaphysical conclusions because *correlation* is not *causation*. Therefore, once a probabilistic model is formulated one should only talk about random variables, their values and their correlations.

As we explained in the section 3, in Bell tests we have 2 local independent random experiments to choose 4 pairs of labels/inputs (x, y) and 4 pairs of correlated distant random experiments corresponding to these 4 choices. These experiments are described by empirical probability distributions of 4 pairs of random variables ($A_{xy}$, $B_{xy}$). Bell-CHSH inequalities cannot be derived and estimated pairwise expectations $E(A_{xy} B_{xy})$ are not constrained by these inequalities.

In order to explain statistical regularities in experimental data one can test different probabilistic models and postulate probabilistic couplings. In this article, random variables in probabilistic models are denoted ($A'_{xy}$, $B'_{xy}$) in order to be not confounded with empirical random variables ($A_{xy}$, $B_{xy}$).

Using quantum coupling (2) one can prove quantum-CHSH inequality: $S \leq 2\sqrt{2}$ [42-44]. Using local realistic (3), local stochastic (5) or LHVM (7-9) couplings one can derive rigorously Bell-CHSH inequalities. Therefore, according to a minority point of view, which I share, the violation of Bell-CHSH proves only that the probabilistic coupling LHVM is inconsistent with the experimental data.

Purely mathematical conditions in LHVM are called *local causality*, *measurement independence* or *freedom- of- choice* etc., what leads to incorrect conclusions and unfounded metaphysical speculations. In particular, as we explained in detail in the sections 4 and 5, closing of *freedom- of- choice loophole* in [7-17] (by choosing randomly inputs (x, y)) does not close the *contextuality loophole* [64, 65]. Namely, if the variables describing measuring instruments/procedures in different measuring settings are correctly incorporated in a probabilistic model, then $P(\lambda | x, y) \neq P(\lambda)$ and of course Bell-CHSH inequalities cannot be derived. Different arguments justifying the violation of *statistical independence* may be found in [53, 54]

The violation of *statistical independence* (9) is sufficient to explain the violation of Bell-CHSH inequalities, but an acceptable contextual hidden variable model has to explain also imperfect

"nonlocal" correlations between the outputs of distant experiments together with their angular dependence. It has also respect EPR–locality: actions or observations in one location do not have immediate effects at other locations [14, 46, 96].

In the section 5, we defined and discussed the contextual model CHVM (10-12).. This model is not Bell-locally-causal, but it is EPR-locally-causal because outputs (a, b) are determined locally by variables $(\lambda_1, \lambda_2)$ describing distant physical systems/qubits and variables $(\mu_x, \mu_y)$ describing measurement contexts (instruments and procedures). Namely $a = A_x(\lambda_1, \mu_x)$ and $b = B_y(\lambda_2, \mu_y)$. The outputs are correlated due to a careful preparation of distant qubits and correlated rotations preceding the local measurements.

The observed sinusoidal oscillations of expectation values $E(A_{xy} B_{xy})$ depend only on the angle $\theta_{xy} = \theta_x - \theta_y$, where $(\theta_x, \theta_y)$ are respective angles by which distant qubits are rotated before local read-outs. This dependence is due to the rotational global symmetry and not due to the spooky influences. A probabilistic models allow only to reproduce statistical regularities observed in the experimental data, but do not provide any information about physical processes and complicated experimental procedures due to which these data were produced. Our contextual model can be represented by a following correlation graph between random variables:

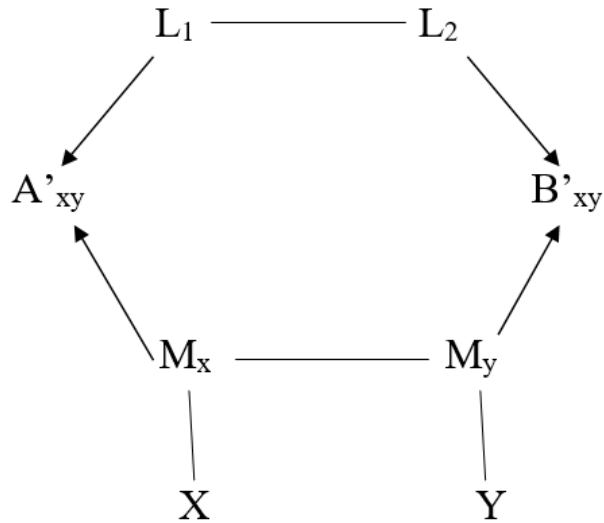

**Figure 1.** Statistically independent random variables X and Y describe choices of inputs. Statistically dependent random variables $L_1$ and $L_2$ describe distant entangled qubits. Statistically dependent random variables $M_x$ and $M_y$ describe correlated local qubit rotations and/or polarizers. Variables $A'_{xy}=A'_x(L_1, M_x)$ and $B'_{xy}=B'_x(L_2, M_y)$ describe correlated outputs (a, b).

The correlation graph should not be confounded with the causal graphs discussed in [93-95]. In (11), joint probabilities $P(\lambda_1, \lambda_2) \neq P(\lambda_1)P(\lambda_2)$ and do not depend on the choice of the inputs. Global space-time symmetries and resulting conservation of : linear momentum, energy and angular momentum are also valid in quantum physics and they are a natural source of "nonlocal correlations" in various experiments and phenomena.

Rotational invariant preparation of distant qubits and global rotational invariance imply that P(a,b|xy) can only depend on $\theta_{xy} = \theta_x - \theta_y$ and not on the individual rotation angles Therefore, in CHVN (10-12) one has to assume that $P(\mu_x, \mu_y | xy) = f(\theta_{xy}, \mu_x, \mu_y)$ or

$P(\mu_x, \mu_y | xy) = P(\mu_x, g(\mu_x, \theta_{xy}) | xy)$ and find a suitable function $f(\theta_{xy}, \mu_x, \mu_y)$ or $g(\mu_x, \theta_{xy})$ in order to explain experimental data.

The model (10-12) may provide only *ad hoc* explanation of "nonlocal correlations" and it is not meant to replace quantum description of physical phenomena underlying Bell Tests, quantum information and quantum computing. CHVN is constructed in order to cut short extraordinary metaphysical speculations based on the violation of Bell-CHSH inequalities in Bell Tests. To understand "nonlocal quantum correlations" one does not need to evoke *quantum nonlocality*, *retro-causality* or *quantum magic*.

The dependence of $P(\mu_x, \mu_y | xy)$ on $\theta_{xy}$ is quite intuitive. For example parameters describing internal structure of two rotated distant Wollaston Prisms, as perceived by two identical and parallel photonic beams, should depend on the difference of the angles formed by Prisms' optical axes. Similarly the parameters describing specific laser pulses or microwaves rotating qubits in distant locations, in various Bell Tests, should also be related due to the rotational invariance. Several arguments against *quantum nonlocality* and in favor of rotational invariance were reviewed recently by Karl Hess [97].

As we mentioned in the introduction, this article is our reply to Nicolas Gisin, who claimed that: *quantum correlations emerge from outside space-time* [24]. We conclude instead, that "*nonlocal quantum correlations*" reported in Bell Tests are due global symmetries of the space-time.

## 10. Appendix

i) <u>Delft experiment</u>

In Delft experiment [10], Alice and Bob own each a single nitrogen vacancy (*NV*) centre: electron spin in diamond. NV centres are prepared using laser excitation pulses and spins are rotated with microwaves. Rotations define the measurement settings (*x*, *y*) and for each setting (*x*, *y*) NV can be in a bright state (+1) in which many photons are emitted or in a dark state (−1) when no photon is emitted.

Using synchronized clocks Alice's and Bob's photon detectors register a click (coded +1) or no click (−1). Each click has its own time-tag. Photons emitted by the NV centres are also sent to overlap at a distant beam splitter C. Each detected photon at C has also its time-tag. Coincidence detection at the output ports of this beam splitter indicates successful photon entanglement and the information of a successful entanglement between the NV centres. This event-ready signal is occurring within a precisely defined window after the arrival time of a sync pulse.

It is an excellent loophole free experiment. Nevertheless, one may not avoid event-ready sampling and several anomalies including an apparent violation of no signalling were reported and discussed [5, 70, 73].

ii) Munich experiments

In Munich experiments [14, 15, 16, 63], a significantly different experimental protocol, based also on entanglement swapping, is used. In [16], two single $^{87}$Rb atoms are placed in two optical dipole traps 400 m apart and connected by a 700 m long optical fiber channel.

The sequence starts by synchronously exciting the single atom in each trap to the state $5^2 P_{3/2} |F'=0, m_{F'}=0\rangle$ which decays to the ground stated $5^2 S_{1/2} |F=1, m_F=\pm 1\rangle$ spontaneously emitting single right- or left-circular polarized photons, $|R\rangle$ or $|L\rangle$ respectively. The photons emitted by distant atoms are guided to an interferometric Bell state measurement setup (BSM) consisting of a fiber beam splitter (BS), polarizing beam splitters (PBS) in each of the output ports, and two detectors in each output port. Their coincident detection heralds the creation of entanglement between distant atoms described by a $|\Psi^+\rangle$ state, if both photons are detected in one output port of the beam splitter with detectors for different polarization, or by $|\Psi^-\rangle$, if one photon is detected in each output port of the beam splitter with detectors for different polarization where:

$$|\Psi^\pm\rangle = \frac{1}{\sqrt{2}}\left(|\uparrow\rangle_x |\downarrow\rangle_x \pm |\downarrow\rangle_x |\uparrow\rangle_x\right) \quad (A1)$$

Given a successful projection, a 'ready' signal is sent to the trap set-ups.

After an additional dephasing and re-phasing of atomic states, the measurement procedure starts with selecting the analysis directions (setting labels (x, y)) according to the output of quantum random number generators (QRNG) or human choices. Then the two atomic qubits are independently analyzed by state-selective ionization and a subsequent detection of the ionization fragments. A particular state of the atomic qubit is ionized and leaves the trap depending on the polarization $\zeta = \cos(\gamma)V + e^{-i\phi}\sin(\gamma)H$ of a read-out laser pulse ($\gamma = \alpha$ for Alice's and $\gamma = \beta$ for Bob's device). If the atom is still in the trap, it is thus projected onto the state

$$|D\rangle = \sin(\gamma)|\downarrow\rangle_x + e^{-i\phi}\cos(\gamma)|\uparrow\rangle_x \tag{A2}$$

(18)

The presence of the atom in a trap is then tested using fluorescence collection at 780 nm, which yields the final measurement outcomes $A_i$ and $B_i$, respectively. The resulting $^{87}Rb^+$ ion and electron are accelerated by electric field to two channel electron multipliers. As the results are reported every time, the detection efficiencies of Alice's and Bob's measurements are effectively one. Any component loss or ionization inefficiency contributes to the noise in the quantum channel [16].

The main topic discussed in [16] was device-independent quantum key distribution system but also the violation of CHSH inequality was reported: S = 2.578(75).

Very detailed discussion of earlier Munich experiments may be found in [63]. The significant violation of inequalities was confirmed but several anomalies were still reported. In the section 5.2.3 we may read: "*All four correlators have significantly different absolute values as it would be expected from the atom-photon state and the settings. This is not critical for a Bell test. More critical anomaly is a correlation of the photon input choice A with the measurement outcome of the atom. This is quite notable since the measurement of the atoms should be independent from the measurement settings for the photon for both local realism as well as quantum mechanics*"

iii) <u>Zurich Experiment</u>

In this experiment [17], two superconducting circuits behaving like transmon-style qubits are housed in two refrigerators A and B separated by 30 m cryogenic link. Their states and transmission frequencies are controlled on nanosecond time scale using amplitude and phase modulated pulses and magnetic flux bias pulses. Resonators combined with Purcell filters are used to read the state of each qubit.

Distant qubits are entangled using of a photon-transfer resonators, which are coupled using a coaxial line to the aluminum waveguide connecting the two sites. Both qubits and their support circuitry are fabricated on two nominally identical chips

In each individual trial of a Bell Test, using direct photon exchange, a Bell state

$$|\Psi^+\rangle = \frac{1}{\sqrt{2}}\left(|g\rangle|e\rangle + |e\rangle|g\rangle\right) \tag{A3}$$

is generated. In (A3), $|g\rangle$ and $|e\rangle$ are respectively ground and excited states of corresponding qubits.

The authors denote inputs (a, b) and the outputs (x, y), we report below their results using the convention used in this article: (x, y) denote inputs and (a, b) outputs. The input bits *x* and *y* for the measurement basis choice are generated independently at each node using random number generators (RNG).

These inputs become available as a voltage pulses at the output of the corresponding RNGs. These pulses control microwave switches that conditionally pass a microwave basis-rotation pulse to the qubit at nodes A (B).

Next as we read in [17]:

*After the microwave pulse has fully rotated both qubit states into the randomly chosen basis, we read out the qubits at A and B by applying a microwave tone to their dedicated readout resonators. We detect the amplitude and phase of the readout pulse after several stages of amplification… record it with a digitizer (analogue-to-digital converter, ADC) and post-process the data with a field programmable gate array… We achieve single-shot readout fidelities of $F_r^A$ = 99.05% and $F_r^B$ = 97.60% in only 50 ns integration time…As done with the random basis choice signals, we route the readout signals through the side ports of the dilution refrigerators at sites A and B.*

The basis choices (x, y) and the corresponding readout result (a, b) are recorded for all n trials. Then taking into account all n trials and closing *the detection loophole* discussed in the section 5, empirical expectation values $E(A_{xy}B_{xy})$ are estimated for all four combinations of measurement basis choices. The authors run n=$2^{20}$ individual trials in each of four consecutive experiments and report maximal violation of CHSH inequality: S=2.0747 ± 0.0033. They also sweep the angle between the two randomly chosen measurement bases observing the expected sinusoidal oscillations.

Please note that experimental data in this experiment allow not only to reject the probabilistic coupling defined by LHVM but they also allow to reject the quantum-coupling (2) using the state (A3), since in spite of excellent readout fidelities S is significantly smaller than $2\sqrt{2}$.

Of course, Quantum- coupling is very flexible and the data can be reasonably fitted using a different density matrix describing prepared entangled qubits.

## References

1. Bohm D., Quantum Theory, New York: Prentice Hall. 1989 reprint, New York: Dover, ISBN 0-486-65969-0
2. Einstein A., Physics and Reality. Journal of the Franklin Institute,1936, 221, 349.
3. Bell, J.S. On the problem of hidden variables in quantum theory. Rev. Mod. Phys. 1966, 38, 450.

4. Clauser J. F., Horne M. A., Shimony A. and Holt R. A., Proposed Experiment to Test, Phys. Rev. Lett. **23**, 880 (1969).
5. Kupczynski M. Is Einsteinian no-signalling violated in Bell tests? Open Physics, 2017, 15 , 739-753, DOI: https://doi.org/10.1515/phys-2017-0087,2017.
6. Kupczynski M., Is the Moon there when nobody looks: Bell inequalities and physical reality, Front. Phys., 23 September 2020 | https://doi.org/10.3389/fphy.2020.00273
7. Aspect A., Grangier P., Roger G.,Experimental test of Bell's inequalities using time-varying analyzers, Phys. Rev. Lett. 1982, 49, 1804-1807.
8. Weihs G., Jennewein T., Simon C., Weinfurther H., Zeilinger A., Violation of Bell's inequality under strict Einstein locality conditions, Phys. Rev. Lett. , 1998, 81, 5039-5043.
9. Christensen B.G., McCusker K.T., Altepeter J.B., Calkins B., LimC.C.W., Gisin N., KwiatP.G., Detection-loophole-free test of quantum nonlocality, and applications, Phys. Rev. Lett., 2013, 111, 130406.
10. Hensen B., Bernien H., Dreau A.E., Reiserer A., Kalb N., Blok M.S. et al., Loopholefree Bell inequality violation using electron spins separated by 1.3 kilometres, Nature, 2015, 15759.
11. Giustina M., Versteegh M.A., Wengerowsky, S., Handsteiner J., Hochrainer A., Phelan K. et al., Significant-loophole-free test of Bell's theorem with entangled photons, Phys. Rev. Lett., 2015, 115, 250401.
12. Shalm L.K., Meyer-Scott E., Christensen B.G., Bierhorst P., Wayne M.A., Stevens M.J. et al., Strong loophole-free test of local realism, Phys. Rev. Lett., 2015, 115, 250402
13. Handsteiner, J. et al. Cosmic Bell test measurement settings from Milky Way stars. *Phys. Rev. Lett.* **2017**, *118*, 060401
14. The BIG Bell Test Collaboration ,Challenging local realism with human choices,The BIG Bell Test Collaboration, Nature, volume 557, pages 212-216 (2018)DOI: 10.1038/s41586-018-0085-3
15. Rosenfeld, W. et al. Event-ready Bell test using entangled atoms simultaneously closing detection and locality loopholes. *Phys. Rev. Lett.* **119**, 010402 (2017).
16. Zhang,W. et al. A device-independent quantum key distribution system for distant users, Nature 607,( 2022)
17. Storz S et.al.: Loophole-free Bell inequality violation with superconducting circuits. Nature, 10 May 2023. external page doi: 10.1038/s41586-023-05885-0
18. Clauser, J.F., Horne, M.A. Experimental consequences of objective local theories. *Phys. Rev. D* **1974**, *10*, 526
19. Ballentine L.E. 1998 *Quantum mechanics: a modern development*. Singapore: World Scientific.
20. Kupczynski M., Seventy years of the EPR paradox, AIP Conf. Proc., 2006, 861, 516-523.
21. Khrennikov, A. *Contextual Approach to Quantum Formalism*; Springer: Dordrecht, The Netherlands, 2009
22. Kupczynski, M., Can we close the Bohr-Einstein quantum debate? Phil.Trans.R.Soc.A., 2017, 20160392., DOI: 10.1098/rsta.2016,0392
23. Khrennikov, A. ,Contextuality, Complementarity, Signaling, and Bell Tests. *Entropy* **2022**, *24*, 1380.
24. Gisin N. Quantum nonlocality: how does nature do it? *Science* **326**, 1357–1358. (2009)
25. Bohr, N. The Philosophical Writings of Niels Bohr. Ox BowPress: Woodbridge, UK, 1987.
26. Khrennikov, A., Can there be given any meaning to contextuality without incompatibility? Int. J. Theor. Phys., 2020;https://doi.org/10.1007/s10773-020-04666-z .'
27. Plotnitsky A. Niels Bohr and Complementarity: An Introduction; Springer: Berlin, Germany; New York, NY, USA, 2012.
28. Kochen S. , Specker E. P., The problem of hidden variables in quantum mechanics. Journal of Mathematics and Mechanics, 17:59–87, 1967.
29. Dzhafarov, E.N., Kujala, J.V. (2014). Contextuality is about identity of random variables.PhysicaScripta T163:014009..
30. Dzhafarov, E.N., Kujala, J.V., Larsson, J.-Å. (2015). Contextuality in three types ofquantum-mechanical systems. Foundations of Physics 7, 762-782.
31. Kujala, J.V., Dzhafarov, E.N., Larsson, J-Å (2015). Necessary and sufficient conditions for extended noncontextuality in a broad class of quantum mechanical systems. Physical ReviewLetters 115:150401.
32. Cervantes, V.H., Dzhafarov, E.N. (2018). Snow Queen is evil and beautiful: Experimental evidence for probabilistic contextuality in human choices. Decision 5:193-204.
33. Dzhafarov, E.N. (2019). On joint distributions, counterfactual values, and hidden variables in understanding contextuality. Philosophical Transactions of the Royal Society A 377:20190144.
34. Kujala, J.V., Dzhafarov, E.N. (2019). Measures of contextuality and noncontextuality. PhilosophicalTransactions of the Royal Society A 377:20190149. (available as arXiv:1903.07170.)
35. Dzhafarov, E.N. (2022). Contents, Contexts, and Basics of Contextuality. In: Wuppuluri, S., Stewart, I. (eds) From Electrons to Elephants and Elections. The Frontiers Collection. Springer, Cham. https://doi.org/10.1007/978-3-030-92192-7_16
36. Kupczynski, M (2021) Contextuality-by-default description of Bell tests: Contextuality as the rule and not as an exception. Entropy **23**(9), 1104 (15pp.) https://doi.org/10.3390/e23091104
37. Araujo M., M. T. Quintino M.T., Budroni C.,Cunha M.T., and Cabello A., All noncontextuality inequalities for the n-cycle scenario. Phys. Rev. A 88, 022118 (2013);https://arxiv.org/pdf/1206.3212.pdf

38. Kupczynski, M. (2023) Contextuality or Nonlocality: What Would John Bell Choose Today? Entropy **25**(2), 280 (13 pp.) https://doi.org/10.3390/e25020280
39. Cetto, A.M., Valdes-Hernandez, A., de la Pena, L. On the spin projection operator and the probabilistic meaning of the bipartite correlation function. *Found. Phys.* **2020**, *50*, 27–39.
40. Bell, J.S. (1964) On the Einstein-Podolsky-Rosen paradox. Physics 1964, 1, 195-200.
41. Bell, J. S., Speakable and Unspeakable in Quantum Mechanics. Cambridge UP, Cambridge (2004)
42. Cirel'son BS. Quantum generalizations of Bell's inequality. *Lett Math Phys.* (1980) **4**:93–100. doi: 10.1007/BF00417500
43. Landau L.J. On the violation of Bell's inequality in quantum theory. *Phys Lett A*. (1987) **20**:54. doi: 10.1016/0375-9601(87)90075-2
44. Khrennikov, A., Two Faced Janus of Quantum Nonlocality, *Entropy* **2020**, *22*(3), 303; https://doi.org/10.3390/e22030303
45. Kupczynski M., Bell Inequalities, Experimental Protocols and Contextuality. Found. Phys., 2015, 45, 735-753.
46. Wiseman, H.M, The two Bell's theorems of John Bell J. Phys. A **47**, 424001 (2014);
47. Wiseman, H.M, Physics: Bell's theorem still reverberates *Nature* volume 510, pages 467–469 (2014)
48. Hall, M.J.W. Local Deterministic Model of Singlet State Correlations Based on Relaxing Measurement Independence. *Phys. Rev. Lett.* **2010**, *105*, 250404.
49. Myrvold, W.; Genovese, M.; Shimony, A. *Bell's Theorem*; Fall 2020 Edition; Edward, N.Z., Ed.; The Stanford Encyclopedia of Philosophy, 2020; Available online: **https://plato.stanford.edu/archives/fall2020/entries/bell-theorem/** (accessed on 10 December 2022).
50. Blasiak, P.; Pothos, E.M.; Yearsley, J.M.; Gallus, C.; Borsuk, E. Violations of locality and free choice are equivalent resources in Bell experiments. *Proc. Natl. Acad. Sci. USA* **2021**, *118*, e2020569118.
51. Kupczynski, M. A comment on: The violations of locality and free choice are equivalent resources in Bell experiments. *arXiv* **2021**, arXiv:2105.14279; https://doi.org/10.48550/arXiv.2105.14279
52. Hance, J.R.; Hossenfelder, S.; Palmer, T.N. Supermeasured: Violating Bell-Statistical Independence without violating physical statistical independence. *Found. Phys.* **2022**, *52*, 81.
53. Hance, J.R.; Hossenfelder, S. Bell's theorem allows local theories of quantum mechanics. *Nat. Phys.* **2022**, *18*, 1382
54. Bell, J.S.; Clauser, J.F.; Horne, M.A.; Shimony, A.A. An Exchange on Local Beables. *Dialectica* **1985**, *39*, 85–96.
55. Bell, J.S, Bertlmann's socks and the nature of reality, CERN preprint Ref.TH.2926- CERN (1980)
56. Bertlmann, R.A. Real or not real that is the question. *Eur. Phys. J. H* **2020**, *45*, 205–236.
57. De Raedt H. et al., Einstein–Podolsky–Rosen–Bohm experiments: A discrete data driven approach, Annals of Physics, Volume 453, June 2023, 169314, **https://doi.org/10.1016/j.aop.2023.169314**
58. Dzhafarov E. N., Assumption-Free Derivation of the Bell-Type Criteria of contextuality/Nonlocality, *Entropy* 2021, *23*(11),1543 43
59. Khrennikov, A. Get rid of nonlocality from quantum physics. *Entropy* **2019**, *21*, 806
60. Kupczynski M., Bertrand's paradox and Bell's inequalities, Phys.Lett. A, 1987, 121, 205-207.
61. Kupczynski M., Closing the Door on Quantum Nonlocality, Entropy, 2018, 20, https://doi.org/10.3390/e20110877
62. Hess K., and Philipp W., Bell's theorem: critique of proofs with and without inequalities. AIP Conf. Proc., 2005, 750, 150-157.
63. Redeker K. PhD thesis, Entanglement of Single Rubidium Atoms: From a Bell Test Towards Applications). https://xqp.physik.uni-muenchen.de/publications/theses_phd/phd_redeker.html
64. Nieuwenhuizen T.M., Is the contextuality loophole fatal for the derivation of Bell inequalities, Found. Phys. 2011, 41, 580-591.
65. Nieuwenhuizen T.M., Kupczynski M., The contextuality loophole is fatal for derivation of Bell inequalities: Reply to a Comment by I. Schmelzer. Found. Phys., 2017, 47, 316-319, DOI: 10.1007/s10701-017-0062-y
66. Larsson J.-A., Loopholes in Bell inequality tests of local realism, J. Phys. A: Math. Theor., 2014, 47, 424003. **DOI**: 10.1088/1751-8113/47/42/424003
67. Kupczynski M., De Raedt H., Breakdown of statistical inference from some random experiments, Comp. Physics Communications, 2016, 200,168.
68. Kupczynski M., On operational approach to entanglement and how to certify it, International Journal of Quantum Information,2016, 14, 1640003.
69. Adenier G., Khrennikov A.Yu., Is the fair sampling assumption supported by EPR experiments?, J .Phys. B: Atom. Mol. Opt. Phys., 2007, 40, 131-141.
70. Adenier G., Khrennikov A.Yu., Test of the no-signaling principle in the Hensen loophole-free CHSH experiment, Fortschritte der Physik , 2017, https://doi.org/10.1002/prop.201600096
71. De Raedt H., Michielsen K., F. Jin, Einstein-Podolsky-Rosen-Bohm laboratory experiments: Data analysis and simulation, AIP Conf. Proc., 2012, 1424, 55-66.

72. De Raedt H., Jin F., Michielsen K., Data analysis of Einstein-Podolsky-Rosen-Bohm laboratory experiments. Proc. of SPIE, 2013, 8832, 88321N1-11.
73. Bednorz A., Analysis of assumptions of recent tests of local realism, Phys. Rev. A, 2017, 95, 042118.
74. Liang, Y.C.; Zhang, Y. Bounding the plausibility of physical theories in a device-independent setting via hypothesis testing. *Entropy* **2019**, *21*, 185.
75. Iannuzzi M., Francini R, Messi R. and Moricciani D.:Bell-type Polarization Experiment With Pairs Of Uncorrelated Optical Photons, Physics Letters A 384 (2020) 126200;Doi:10.1016/j.physleta.2019.126200 (arXiv:2002.02723 [quant-ph])
76. Larsson, J.-A. and Gill R.D., Bell's inequality and the coincidence-time loophole. Europhys. Lett., 2004, 67, 707-13
77. Pearle P., Hidden-Variable Example Based upon Data Rejection, Phys. Rev. D 2, 1418, 1970, DOI:https://doi.org/10.1103/PhysRevD.2.1418
78. Garg, A. & Mermin, N. D. Detector inefficiencies in the Einstein-Podolsky-Rosen experiment. *Phys. Rev. D* **35**, 3831 (1987)
79. J.-Å. Larsson (1998), Bell's inequality and detector inefficiency. Phys. Rev. A 57 3304–8.
80. N. Gisin and B. Gisin (1999), A local hidden variable model of quantum correlation exploiting the detection loophole. Phys. Lett. A 260, 323–327
81. De Raedt H, Michielsen K, Hess K. The photon identification loophole in EPRB experiments: computer models with single-wing selection. Open Phys (2017) 15:713–33.doi:10.1515/phys-2017-0085
82. Kupczynski M (2023), Response:"Commentary: Is the moon there if nobody looks? Bell inequalities and physical reality".Front. Phys. 11:1117843.doi: 10.3389/fphy.2023.1117843
83. Cirac, J. I., Zoller, P., Kimble, H. J. & Mabuchi, H. Quantum state transfer and entanglement distribution among distant nodes ina quantum network.Phys. Rev. Lett. 78, 3221 (1997); DOI :https://doi.org/10.1103/PhysRevLett.78.3221-y
84. Zangi SM, Shukla C, Ur Rahman A, Zheng B. Entanglement Swapping and Swapped Entanglement. Entropy (Basel). 2023 Feb 25;25(3):415. doi: 10.3390/e25030415. PMID: 36981304; PMCID: PMC10047960
85. Salimian, S., Tavassoly, M.K. & Ghasemi, M. Multistage entanglement swapping using superconducting qubits in the absence and presence of dissipative environment without Bell state measurement. Sci Rep 13, 16342 (2023). https://doi.org/10.1038/s41598-023-43592
86. Khrennikov, A. *Ubiquitous Quantum Structure*; Springer: Berlin, Germany, 2010
87. Basieva, I., Cervantes, V.H., Dzhafarov, E.N., Khrennikov, A. (2019). True contextuality beats direct influences in human decision making. Journal of Experimental Psychology: General 148,1925-1937.
88. Cervantes, V.H., Dzhafarov, E.N. (2019). True contextuality in a psychophysical experiment. Journal of Mathematical Psychology 91, 119-127.
89. Aerts, D., AertsArgu¨elles, J., Beltran, L., Geriente, S., Sassoli de Bianchi, M., Sozzo, S &Veloz, T. (2019). Quantum entanglement in physical and cognitive systems: a conceptual analysis and a general representation. European Physical Journal Plus 134: 493, doi: 10.1140/epjp/i2019-12987-0.
90. Aerts, D., Arguëlles, J.A., Beltran, L., Geriente, S., Sozzo, S. (2023). Entanglement in Cognition Violating Bell Inequalities Beyond Cirel'son's Bound. In: Plotnitsky, A., Haven, E. (eds) The Quantum-Like Revolution. Springer, Cham. https://doi.org/10.1007/978-3-031-12986-5_15
91. Aerts D., The physical origin of the Einstein-Podolsky-Rosen paradox and how to violate Bell inequalities by macroscopical systems, Symposium on the Foundations of Modern Physics: 50 Years of the Einstein-Podolsky-Rosen Gedanken Experiment : Proceedings, Lahti P. and Mittelstaedt P. eds (World Scientific, Singapore, 1985).
92. Mermin, N. D. Hidden variables and the two theorems of John Bell. Rev. Mod. Phys. 65, 803–815 (1993).
93. Wood C.J. and Spekkens R.W, The lesson of causal discovery algorithms for quantum correlations: Causal explanations of Bell-inequality violations require fine-tuningNew Journal of Physics 17, 033002 (2015)
94. Wiseman H.M. and Cavalcanti E.G., "Causarum Investigatio and the Two Bell's Theorems of John Bell," in Quantum [Un]Speakables II: Half a Century of Bell's Theorem, edited by R. Bertlmann and A. Zeilinger (Springer International Publishing, Cham, 2017) pp. 119–142.
95. Pearl J.C. and Cavalcanti,E.G, Classical causal models cannot faithfully explain Bell nonlocality or Kochen-Specker contextuality in arbitrary scenarios, Quantum 5, 518 (2021) ;https://doi.org/10.22331/q-2021-08-05-518
96. A. Einstein, B. Podolsky, N. Rosen, Can quantum-mechanical description of physical reality be considered complete?, Phys. Rev. 47, 777 (1935)
97. Hess, K.: A Critical Review of Works Pertinent to the Einstein-Bohr Debate and Bell's Theorem, *Symmetry* **2022**, *14*(1), 163, **https://doi.org/10.3390/sym14010163**